\documentclass[aps,preprint,amsmath,amssymb,t1,11pt]{revtex4}
\usepackage{graphicx}
\usepackage{epstopdf}
\usepackage{xcolor}
\usepackage{slashed}
\usepackage{multirow}

\begin{document}
\draft

\title{Constraints on $\tau$  electromagnetic moments via tau pair production at the Muon colliders}

\author{H. Denizli}
\email[]{denizli_h@ibu.edu.tr}
\affiliation{Department of Physics, Bolu Abant Izzet Baysal University, 14280, Bolu, T\"{u}rkiye.}

\author{A. Senol}
\email[]{senol_a@ibu.edu.tr}
\affiliation{Department of Physics, Bolu Abant Izzet Baysal University, 14280, Bolu, T\"{u}rkiye.}

\author{M. K{\"o}ksal}
\email[]{mkoksal@cumhuriyet.edu.tr}
\affiliation{Department of Physics, Sivas Cumhuriyet University, 58140, Sivas, T\"{u}rkiye.}
\affiliation{Sivas Cumhuriyet University Nanophotonics Research and Application Center (CUNAM), Sivas, 58140, T\"{u}rkiye.}

\date{\today}

\begin{abstract}

The anomalous magnetic ($\tilde{a}_\tau$) and  electric dipole ($\tilde{d}_\tau$) moment of tau lepton described in $\tau\bar{\tau}\gamma$ vertices are studied via $\mu^{+}\mu^{-}  \rightarrow \tau^+\bar{\tau}^-$ process at the Muon colliders designed with the proposed center-of-mass-energy / integrated luminosity configurations of 3 TeV/ 1 ab$^{-1}$ and 10 TeV/ 10 ab$^{-1}$.
We obtained the  $95\%$ confidence level limits on the $\tilde{a}_\tau$ and $\tilde{d}_\tau$ parameters without and with systematic uncertainty of $10\%$ and compared with the experimental results.
The most stringent limits on the anomalous couplings without systematic uncertainty are $-2.63\times10^{-4}< \tilde{a}_\tau <2. 65\times10^{-4}$ and $\mid \tilde{d}_\tau \mid \leq 1.47\times10^{-18} e cm$ at $\sqrt s =10$ TeV and $L_{int}$=10 ab$^{-1}$ option of the Muon collider. Our results show that the Muon colliders lead to a remarkable improvement on the current experimental limits of the anomalous magnetic and electric dipole moments of the tau lepton.
\end{abstract}

\maketitle

\section{Introduction}

The magnetic dipole moment of the electron, which is responsible for its interaction with the magnetic field in the Born approximation, is expressed as follows:
\begin{equation}
    \vec{\mu} = \Big(\frac{g \mu_B}{\hbar}\Big)\vec{s}\,.
\end{equation}
Here, $g$ is the gyromagnetic factor, $\mu_B$ is the Bohr magneton, and $\vec{s}$ is the spin of the electron. In the Dirac equation, the value of $g$ for the electron is 2. It is customary to describe the deviation of $g$ from 2 in terms of the value of the anomalous magnetic moment
\begin{equation}
    a_e = \frac{g-2}{2}\,.
\end{equation}
However, it was initially determined through Quantum ElectroDynamics (QED) using radiative corrections by Schwinger as $a_e=\alpha_{EM} /(2\pi)\approx 0.00116$ \cite{1}. The precision of $a_e$ has been extensively examined in numerous studies. These investigations have yielded the most accurate determination of the fine-structure constant $\alpha_{QED}$, since $a_e$ is quite senseless to the strong and weak interactions. On the other hand, the precision measurement on muon’s magnetic dipole moment ($a_\mu$) is an important probe to test for the Standard Model (SM) and investigate the new physics beyond the SM 
due to the contribution of mass suppression terms from interactions with massive particles beyond the SM. Therefore, the muon is $(m_\mu/m_e)^2\approx4\times10^{4}$ times more sensitive to such terms with respect to the electron \cite{2}.

There is an increasing interest in the precise measurement of muon's magnetic dipole moment due to the difference between experimental data and the SM prediction, up to $4.2\  \sigma$ \cite{3,4,5}. Currently, the discrepancy between theoretical and observed values at the FNAL Muon $g-2$ experiment is given as follows~\cite{5}
\begin{eqnarray}
    a_\mu^{SM}& =& 116591810(43)\times10^{-11},\\  
a_\mu^{obs}& =&116 592 061(54)\times10^{-11},\\
    a_\mu^{obs}-a_\mu^{SM}&=&(251\pm97)\times10^{-11}.
\end{eqnarray}
The SM contribution to the $a_\tau$ is determined by the summation of the Quantum Electrodynamics (QED), Hadronic (HAD) and Electroweak (EW) terms:
\begin{eqnarray}
a_\tau^{QED}&=&117324\times10^{-8},
\\
a_\tau^{HAD}&=&350.1\times10^{-8},
\\
a_\tau^{EW}&=&47.4\times10^{-8}.
\end{eqnarray}
Combining these contributions leads to the anomalous magnetic moment $a_\tau=117721\times10^{-8}$. Similarly, a precise calculation of the anomalous magnetic moment of tau, the most massive lepton, is thought to provide an excellent opportunity to reveal the implications of new physics beyond the SM. The anomalous magnetic moments of the electron and muon have been studied with high sensitivity using a spin precision experiment. On the other hand, the spin precision experiment is unsuitable for investigating the anomalous magnetic moment $a_\tau$ due to the relatively short lifetime of the tau lepton, which is $2.9\times10^{-13}$ s. For the tau lepton, the measurement of anomalous magnetic moment is investigated by comparing the SM cross section with the cross section measured from tau pair production processes at colliders with high centre-of-mass energies instead of spin precision experiments.

In the SM framework, there are three possible sources of CP violation. One arises from complex couplings in the CKM matrix of the quark sector.
Neutrinos in the SM are massless.  If the SM is corrected to include neutrino masses, CP violation can occur in the mixing of leptons. The third source of CP violation is possible in flavour-conserving strong interaction processes. On the other hand, presence of CP violation in the SM is not sufficient to explain the observed baryon asymmetry of the universe. Therefore, it is evident that there needs to be CP violation beyond the SM. A direct study of CP violation is possible by studying the electric dipole moment of the tau lepton \cite{cp1,cp2}. However, CP violation in the quark sector induces a small electric dipole moment of the tau lepton. Therefore, one has to go to at least three loop levels to obtain a non-zero contribution. 
A theoretical estimate of this value is given by \cite{cp3}
\begin{equation}
    \mid d_\tau \mid \leq 10^{-34}\ e \\ cm.
\end{equation}
Even though the electric dipole moment of the tau lepton is small within the SM, it can cause a detectable size due to interactions coming from physics beyond the SM \cite{cp4,cp5,cp6,cp7,cp8,cp9, cp10, cp11, cp12}.

$a_\tau$ magnetic and $d_\tau$ electric dipole moments of the tau lepton analyses the structure of the interaction of the tau lepton with a photon.
The most general anomalous vertex function that describes the $\tau \bar{\tau} \gamma$ interaction can be parameterised as follows \cite{kos}

\begin{equation}\label{eq:form_factors}
    \Gamma^\nu = F_1(q^2)\gamma^\nu +F_2(q^2)\frac{i\sigma^{\nu\mu}q_\mu}{2m_{\tau}}+F_3(q^2)\frac{\sigma^{\nu\mu}q_\mu\gamma^5}{2m_{\tau}}.
\end{equation}

\noindent Here, $\sigma^{\nu\mu}=i(\gamma^{\nu}\gamma^{\mu}-\gamma^{\mu}\gamma^{\nu})/2$, $q$ is the momentum transfer to the photon and $m_\tau$ is the tau’s mass.
$F_1(q^2)$ and $F_2(q^2)$ are the Dirac and Pauli form factors, $F_3(q^2)$ is the electric dipole form factor. 
The last term $\sigma^{\nu\mu}\gamma^{5}
$ breaks the CP symmetry, so the coefficient $F_3(q^2)$ determines the strength of a possible CP violation process, which could come from new physics beyond the SM. 
$F_1(q^2)$, $F_2(q^2)$ and $F_3(q^2)$ are factors in the limit $q^2\rightarrow0$ correspond to the following formulae
\begin{equation}
 F_1(0)=1,\ \ \ \ \, F_2(0)=a_\tau,\ \ \ \ \ F_3(0)=\frac{2m_{\tau}}{e} d_{\tau}.
\end{equation}

The $68\%$ Confidence Level (C.L.) experimental limit on the anomalous $a_\tau$ coupling of the tau lepton is $-0.088<a_\tau<0.056$, obtained by CMS Collaboration at the LHC via the process $Pb Pb\rightarrow Pb\gamma^* \gamma^* Pb \rightarrow Pb  \tau \bar \tau Pb$ \cite{CMS}.
The DELPHI collaboration at LEP also reports 95\% C.L.limit on $a_\tau$ through the process $e^+e^-\rightarrow e^+\gamma^* \gamma^* e^- \rightarrow e^+ \tau \bar \tau e^- $ at center-of-mass-energy between 183 and 208 GeV with a total integrated luminosity of 650 pb$^{-1}$ as $-0.052<a_\tau<0.013$ \cite{lep1}. 
However, the most restrictive limit at $95\%$
C. L. via the process $pp\rightarrow p\gamma^* \gamma^* p \rightarrow p\tau \bar \tau p $ by CMS Collaboration at the LHC is $-0.0009<a_\tau<0.0009$ \cite{CMS1}.

The experimental limits on $d_\tau$ coupling at $95\%$ C.L. by BELLE Collaboration in the KEKB collider with an integrated luminosity of 29.5 fb$^{-1}$ at $\sqrt{s} = 10.58$ GeV are \cite{lep2}
\begin{eqnarray}
    -2.2  &\leq Re(d_\tau) \leq &4.5 \ \ (\times  10^{-17} \ \ e cm),\\
    -2.5 &\leq  \text{Im}(d_\tau) \leq & 0.8 \ \ (\times  10^{-17} \ \ e cm)
\end{eqnarray}
and later by CMS Collaboration \cite{CMS1} 
\begin{eqnarray}
    -2.9  &\leq d_\tau \leq &2.9 \ \ (\times  10^{-17} \ \ e cm).
\end{eqnarray}

The experiments focused on various high-precision measurements of  tau anomalous magnetic moment via tau pair production in high-energy processes involve off-shell photons or taus in the $\tau \bar{\tau} \gamma$ vertices but the measured quantity is not directly $a_\tau$.
In this case the potential discrepancy between the SM predictions and the observed values of $\tau \bar{\tau} \gamma$ couplings can be investigated through the effective Lagrangian approach in a model-independent manner. This is achieved by considering the following dimension-six effective operator \cite{lag}:
\begin{eqnarray}
L_{eff}=\frac{1}{\Lambda^{2}} [C_{LW}^{33} (\bar{\ell_{\tau}}\sigma^{\mu\nu}\tau_{R})\sigma^{I}\varphi W_{\mu\nu}^{I}+C_{LB}^{33} (\bar{\ell_{\tau}}\sigma^{\mu\nu}\tau_{R})\varphi B_{\mu\nu}+h.c.].
\end{eqnarray}
After the electroweak symmetry breaking, contributions to the magnetic and electric dipole moments of the tau lepton are given by
\begin{eqnarray}
\kappa=\frac{2 m_{\tau}}{e} \frac{\sqrt{2}\upsilon}{\Lambda^{2}} Re[\cos\theta _{W} C_{LB}^{33}- \sin\theta _{W} C_{LW}^{33}],
\end{eqnarray}
\begin{eqnarray}
\tilde{\kappa}=-\frac{\sqrt{2}\upsilon}{\Lambda^{2}} Im[\cos\theta _{W} C_{LB}^{33}- \sin\theta _{W} C_{LW}^{33}].
\end{eqnarray}
Here, $\upsilon$ represents the vacuum expectation value and $\sin\theta _{W}$ shows the weak mixing angle.
The relations between $\kappa$ and $\tilde{\kappa}$ parameters with $\tilde{a}_\tau$ and $\tilde{d}_\tau$ are defined by
\begin{eqnarray}
\kappa=\tilde{a}_\tau, \;\;\;\; \tilde{\kappa}=\frac{2m_\tau}{e}\tilde{d}_\tau.
\end{eqnarray}
Consequently, the magnetic and electric dipole moments of the tau lepton allow stringent tests for new physics beyond the SM and have been studied in detail by Refs. \cite{50,51,52,53,54,55,56,57,58,59,60,61,62,63,64,65,66,67}.

Recently, the study of the physics capabilities of a potential muon collider operating at high energies in the multi-TeV range has attracted considerable interest. The first proposal for such a collider appeared in the European Strategy for Particle Physics in 2020 \cite{mu1,mu2,mu3}. A muon collider with multi-TeV capabilities could facilitate the direct study of new particles over a wide range of unexplored masses. It could also serve as a tool for exploring different models of new physics \cite{mu4}. 
The most likely centre-of-mass energies and luminosities for this collider are proposed to be a centre-of-mass energy of 10 TeV with an integrated luminosity 10 ab$^{-1}$. 
Electroweak vector boson fusion/scattering, the dominant production process in muon collisions at energies above a few TeV, would provide a precise probe for the emergence of new physics processes \cite{mu5,mu6,mu7,mu8}. 
A muon collider at 10 TeV or more would have the potential not only to discover new particles of currently inaccessible mass, including WIMP dark matter candidates, but also to probe cracks in the SM through the sensitive study of the Higgs boson, including the direct observation of the double Higgs production and the precision measurement of the triple Higgs coupling. By combining precision with energy, it will uniquely track the quantum signature of new phenomena in new observables.

This study focuses on the effects of the anomalous $\tau \bar{\tau} \gamma$ interaction through the process $\mu^+\mu^- \to \tau^+\tau^-$ at the future Muon Collider under consideration. The outline of this work as follow: Section II comprises a discussion of the signal process at the cross-section level, accompanied by details of the event generation including detector effects for the signal and relevant SM backgrounds at the Muon Colliders. In Section III, we present the statistical method employed to derive constraints on the anomalous $\tilde{a}_\tau$ and $\tilde{d}_\tau$ couplings of the tau lepton. Finally, we conclude with a summary of the obtained limits on the anomalous $\tilde{a}_\tau$ and $\tilde{d}_\tau$ couplings.

\section{Generation of signal and background events} 
In this study, we focus on the sensitivities of the anomalous magnetic and electric dipole moments describing the anomalous $\tau\bar{\tau}\gamma$ couplings  via $\mu^+\mu^-\to\tau^-\tau^+$ signal process at Muon colliders.
The primary decay channels of the tau lepton, which decays both to lighter leptons such as electrons and muons and to lighter hadrons such as $\pi$'s and $K$'s with a lifetime of ($290.3 \pm 0.5 \times 10^{-15}$ s), can be as a divided into one charged particle (one-prong decay) and three charged particle (three prong decay):
\begin{eqnarray}
\text{One-prong decays:}&&\nonumber\\
&\tau^{\pm}&\to\nu_{\tau}+l^{\pm}+\bar{\nu_l}~~(l=e,\mu),\\
&\tau^{\pm}&\to\nu_{\tau}+\pi^{\pm}~~(\pi~~ \text{mode}),\\
&\tau^{\pm}&\to\nu_{\tau}\rho^-\to\nu_{\tau}\pi^-\pi^0 ~~(\rho~~ \text{mode}),\\
&\tau^{\pm}&\to\nu_{\tau} a_1^-\to\nu_{\tau}\pi^0\rho^-\to\nu_{\tau}\pi^0\pi^0\pi^-  ~~(a_1~\text{mode}),\\
\text{Three-prong decays:}&&\nonumber\\
&\tau^{\pm}&\to\nu_{\tau} a_1^-\to\nu_{\tau}\pi^0\rho^-\to\nu_{\tau}\pi^0\pi^0\pi^-  ~~(a_1~\text{mode}).
\end{eqnarray}
In our analysis, we contemplate the hadronic decay of oppositely charged tau leptons ($\tau^+$ and $\tau^-$). Among the hadronic tau lepton decay modes, we consider the specific one-prong decay processes, namely $\tau^{\pm}\to\pi^{\pm}\nu~~(\pi~~ \text{mode})$, to minimize the potential loss of kinematic information stemming from the presence of multiple undetected neutrinos.The effective Lagrangian that describes $\tau-\nu_{\tau}-\pi$ vertex is presented as follows in Ref. \cite{Hagiwara:2012vz}:
\begin{eqnarray}\label{eq3}
\mathcal{L}_{\pi}= \sqrt{2}G_Ff_1\bar{\tau}\gamma^{\mu}P_L\nu_{\tau}\partial_{\mu}\pi^- +h.c.
\end{eqnarray}
where $G_F$ is Fermi constant, $P_L$ is the chiral projection operator and $f_1$ is the constant form factor as defined $f_1 = f_{\pi}\cos\theta_C$ with Cabibbo angle $\theta_C$.

\begin{figure}
\includegraphics[scale=0.62]{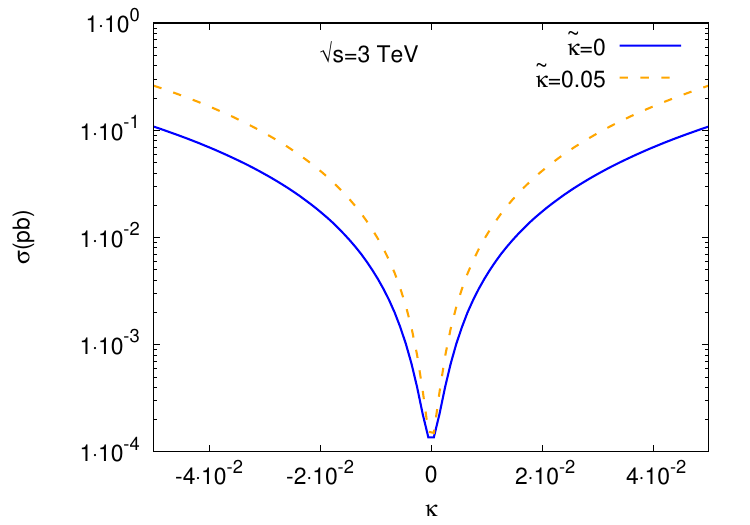}
\includegraphics[scale=0.62]{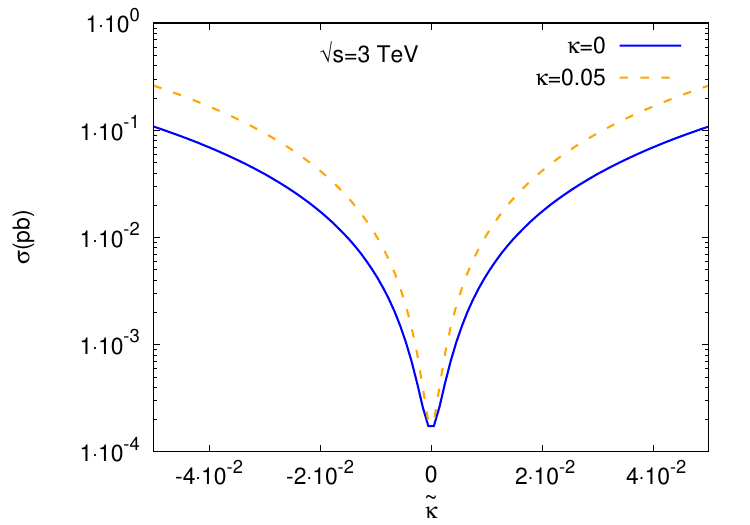}
\caption{The total cross sections
of the process $\mu^{+}\mu^{-}  \rightarrow \tau^+\bar{\tau}^-$ as a
function of $\kappa$ (left panel) and $\tilde\kappa$ (right panel) parameters for the Muon collider at center-of-mass energies of $\sqrt s = 3$ TeV option. \label{cs3TeV}}
\end{figure}

The Monte Carlo events for the signal process $\mu^+\mu^- \to\tau^{+}(\to\pi^{+}\bar\nu_{\tau} )\tau^{-} (\to\pi^{-}\nu_{\tau}$) and relevant background process, having the same final state as the considered signal process including only SM contribution, are generated by using {MadGraph5\_aMC$@$NLO v3\_1\_1} \cite{Alwall:2014hca}. The {\sc Universal FeynRules Output} (UFO) \cite{Alloul:2013bka} model file implementing the $\tau \bar{\tau} \gamma$ vertices given in Eq.(\ref{eq:form_factors}) with {\sc FeynRules} package \cite{Degrande:2011ua} and the hadronic decays of the $\tau^{\pm}$ considered with the Tau Decay model \cite{Hagiwara:2012vz} including the effective Lagrangian given in Eq.(\ref{eq3}) are inserted to {MadGraph5\_aMC$@$NLO v3\_1\_1}. 
In order to observe the effect of the anomalous electric and magnetic moment of tau on the cross-section level, the total cross-sections is evaluated as a function of anomalous parameters $\kappa$ and $\tilde\kappa$, respectively. A strong dependence of the total cross-section with respect to the anomalous parameters $\kappa$ and $\tilde\kappa$ is observed as in Fig.~\ref{cs3TeV}. Furthermore, the deviation from  the SM cross section (varying one coupling  $\tilde\kappa$ ($\kappa$) at a time; the other set to $\kappa$ ($\tilde\kappa$)= 0  as blue line and 0.05 as orange dashed line) ranges about three orders for the interval of $\kappa$ and $\tilde\kappa$ considered in Fig.~\ref{cs3TeV}.
For further analysis including the parton showering and fragmentation as well as the detector response, the generated signal and background events passed through Pythia 8.2 \cite{Sjostrand:2014zea} and Delphes 3.4.2 \cite{deFavereau:2013fsa} with the detector card parameterized for the Muon Collider at the target performance. Jets are clustered by the Valencia Linear Collider (VLC) algorithm \cite{Boronat:2014hva} in inclusive mode with $R = 0.5$  cone size parameter and $\gamma=\beta=1$ using FastJet \cite{Cacciari:2011ma}. Candidate $\tau^{+}(\to\pi^{+}\bar\nu_{\tau}) \tau^{-} (\to\pi^{-}\nu_{\tau})$ events are selected by requiring the presence of high missing transverse momentum and two jets obtained from subsequent decay of taus, tagged as tau-tagged jets ($\tau_{had}$). 
The $\tau_{had}$'s are tagged by using a $\tau$ identification efficiency of about 80\% for $p_T>10$ GeV and have a fake rate about 0.2\% in the Muon collider detector card. The tau-tagged jets with the highest transverse momentum ($p_T$) is labeled as $\tau_1$ and the other one with lower as $\tau_2$. We select a region in phase space where the transverse momentum and the pseudo-rapidity of the tau-tagged jets are $p_T^{\tau_1,\tau_2}\geq 20$ GeV and $|\eta^{\tau_1,\tau_2}|\leq 2.5$, respectively. We also require that the two hadronic tau $\tau_{1}$ and $\tau_{2}$ candidates are oppositely charged. 
Additional cuts on the differences of scattering angle ($\Delta\theta=\theta_{\tau_1}-\theta_{\tau_2}$) and the azimuthal angle ($\Delta\phi=\phi_{\tau_1}-\phi_{\tau_2}$) between two tau candidates are $|\Delta\theta|>$2.0 and $|\Delta\phi|>$2.7, respectively.
 The total transverse mass, also used in ATLAS analyses \cite{ATLAS:2020zms}, is considered as discriminating parameter to separate signal and background events and defined as 
\begin{eqnarray} \label{eq:mTtot}
m_T^{\rm tot} = \sqrt{m_T^2 (\vec{p}_T^{\; \tau_1}, \vec{p}_T^{\; \tau_2}) + m_T^2 (\vec{p}_T^{\; \tau_1}, \vec{p}_T^{\; \rm miss}) + m_T^2 (\vec{p}_T^{\; \tau_2}, \vec{p}_T^{\; \rm miss}) } \,,
\end{eqnarray}	
where each $m_T$ is the  transverse mass between two transverse momenta $p_{T,i}$ and $p_{T,j}$ as given below
\begin{eqnarray} \label{eq:mT}
m_T (\vec{p}_T^{\; i}, \vec{p}_T^{\; j} ) = \sqrt{2 \hspace{0.25mm} p_{T, i} \hspace{0.5mm} p_{T, j} \left ( 1 - \cos \Delta \phi \right )} \,.
\end{eqnarray}

\begin{figure}
\includegraphics[scale=0.45]{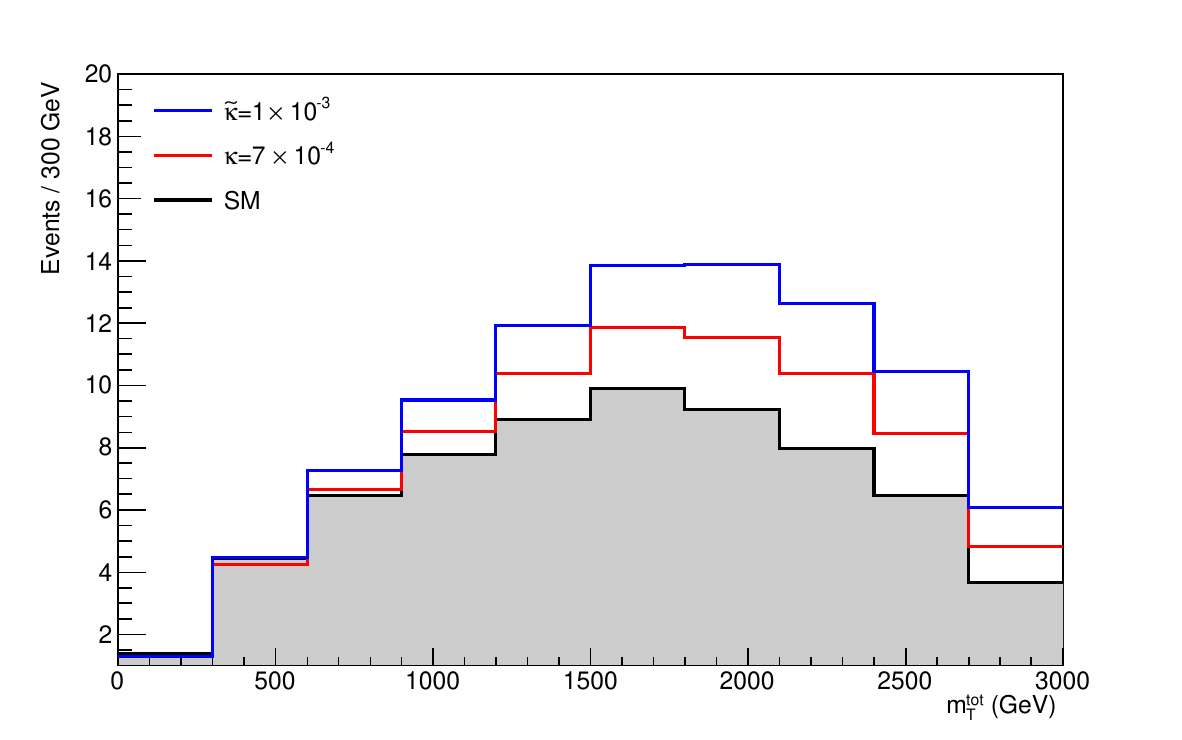}\includegraphics[scale=0.45]{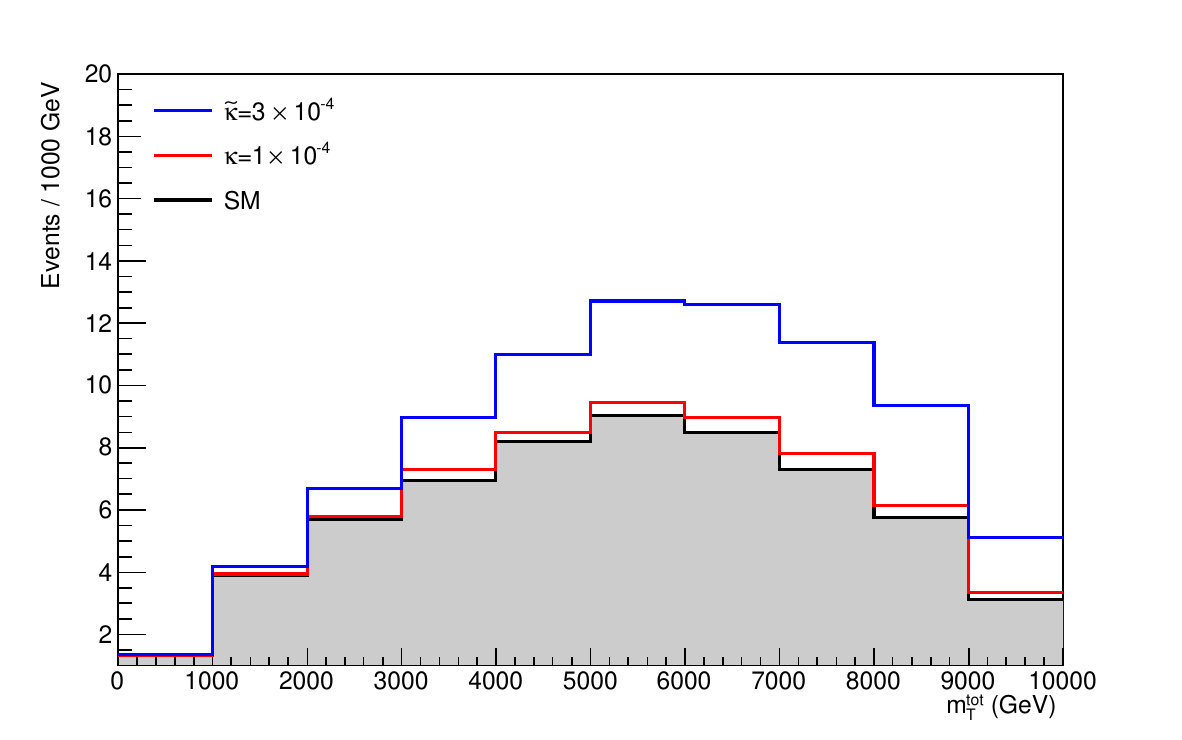}
\caption{Normalized total transverse mass (m$^{tot}_T$) distribution for $\sqrt{s}=3~(10)$ TeV Muon collider at $L_{int}$= 1 (10)~ab$^{-1}$ on left (right) panel. The solid blue, red lines and gray areas show the expectations of a signal due to $\tilde{\kappa} = 1\times10^{-3}$, $\kappa = 7\times10^{-4}$($\tilde{\kappa} = 3\times10^{-4}$, $\kappa = 1\times10^{-4}$) and SM background, respectively. \label{mtransverse}}
\end{figure}

In Fig.\ref{mtransverse}, we show the normalized distributions of $m_T^{\rm tot}$ for $\mu^{+}\mu^{-} \rightarrow \tau^{+}(\to\pi^{+}\bar\nu_{\tau}) \tau^{-} (\to\pi^{-}\nu_{\tau})$ signal and background processes at $\sqrt{s}=3~(10)$ TeV muon collider with $L_{int}$= 1 (10)~ab$^{-1}$ on the left (right) panel. The black lines correspond to the SM background while the blue and red lines correspond to $\tilde\kappa = 1\times10^{-3}$ and $\kappa = 7\times10^{-4}$ for the anomalous parameters of the tau lepton, respectively. 
Deviation between signal with non-zero values of the anomalous parameters ( $\tilde\kappa=\kappa$= $\mathcal{O}(10^{-3})$) and SM background is more clearly seen at the higher mass region of $m_T^{\rm tot}$ distribution in Fig.\ref{mtransverse}.
\section{Sensitivity of anomalous moments of the tau lepton}
\begin{figure}
\includegraphics[scale=0.4]{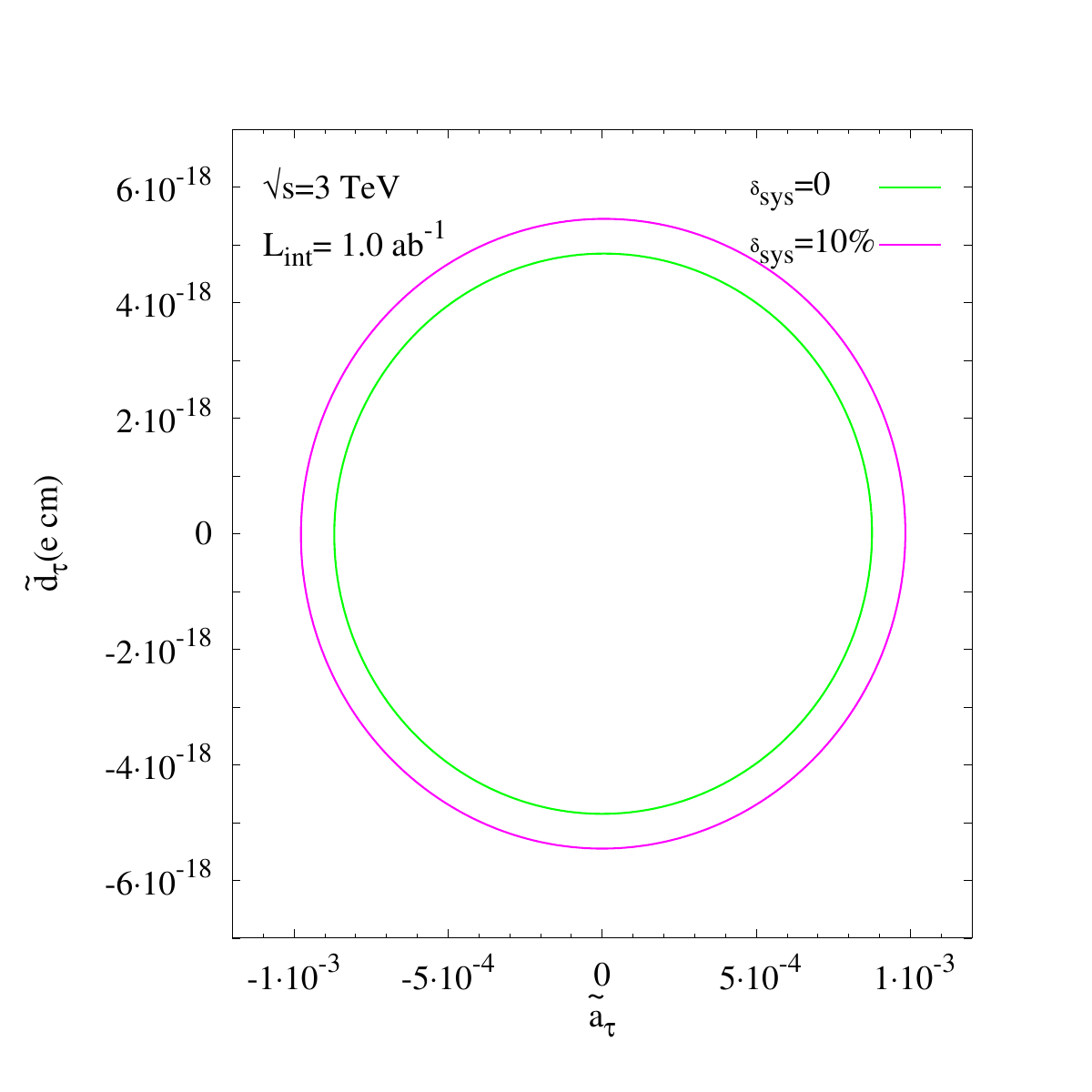}
\includegraphics[scale=0.4]{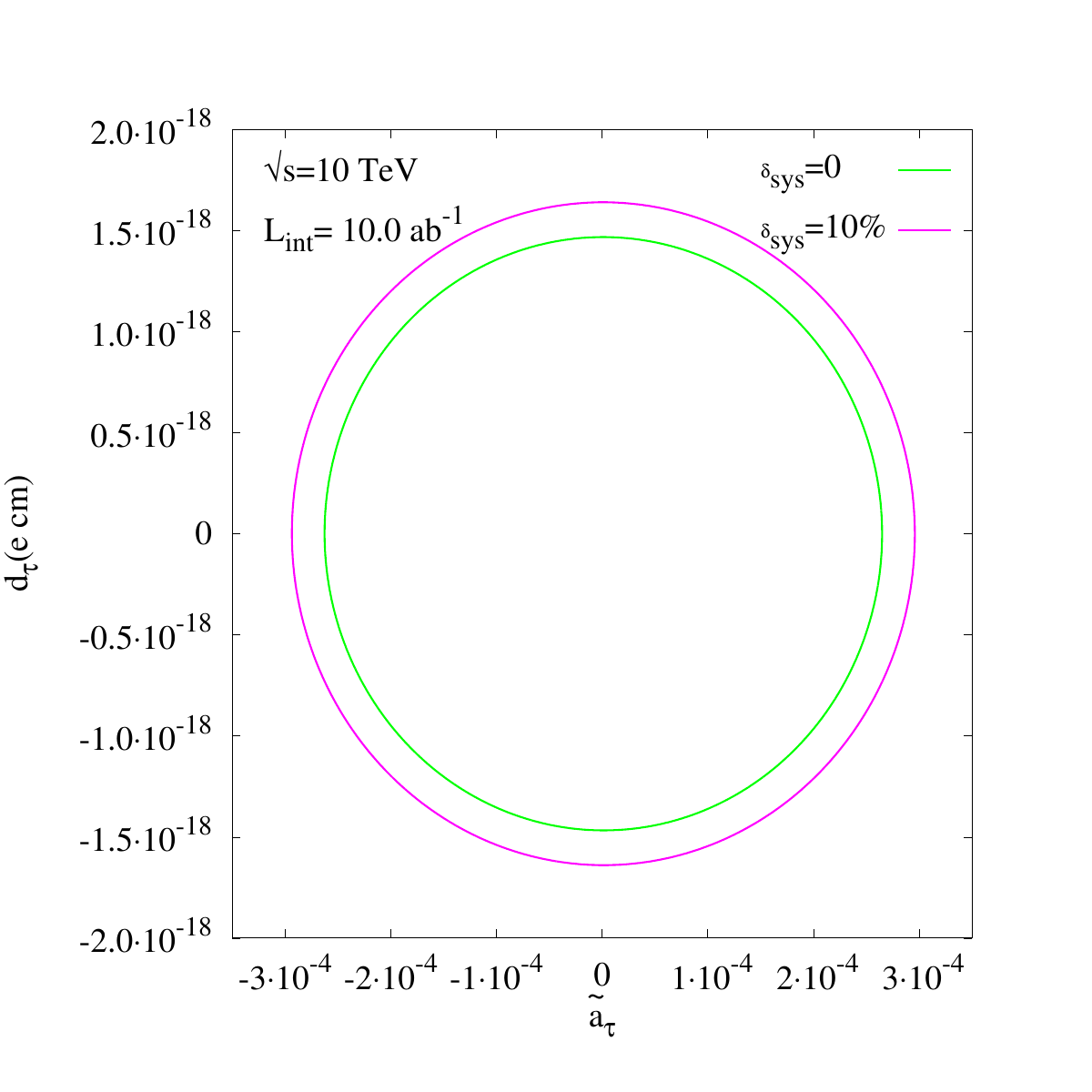}
\caption{Two-dimensional 95\% C.L. intervals on the $\tilde{a}_\tau$ and $\tilde{d}_\tau$ planes considering without and with systematic uncertainty of 10 \%  by the process $\mu^+\mu^- \to\tau^{+}(\to\pi^{+}\bar\nu_{\tau} )\tau^{-} (\to\pi^{-}\nu_{\tau}$) 
for $\sqrt s = 3$ TeV (10 TeV) option $L_{int}$= 1 (10)~ab$^{-1}$ on the left (right) panel. \label{contour_3T}}
\end{figure}

This section is dedicated to obtain sensitivities of the anomalous electric ($\tilde{a}_\tau$) and magnetic ($\tilde{d}_\tau$) dipole moment couplings of the tau lepton by the $\chi^2$ criterion method with systematic uncertainty which is defined by
\begin{eqnarray}\label{eq6}
\chi^{2} (\tilde a_\tau ,\tilde d_\tau)=\sum_i^{n_{bins}}\left(\frac{N_{i}^{NP}(\tilde a_\tau ,\tilde d_\tau)-N_{i}^B}{N_{i}^B\Delta_i}\right)^{2}
\end{eqnarray}
where $N^{NP}$ is the total number of events in the presence of effective couplings $(S)$ and corresponding SM backgrounds $(B)$, $N^B$ is the number of events only coming from relevant SM backgrounds in $i$th bin of normalized $m_T^{\rm tot}$ distribution and $\Delta_i=\sqrt{\delta_{sys}^2+\frac{1}{N_i^B}}$ is the combined systematic ($\delta_{sys}$) and statistical uncertainty  in each bin. In this study, we concentrate on obtaining $95\%$
C. L. limits on $\tilde{a}_\tau$ and $\tilde{d}_\tau$ couplings via the $\mu^+\mu^- \to \tau^+\tau^-$ signal process at the Muon collider with the center-of-mass energies at $\sqrt{s}$= 3~TeV and 10~TeV and the integrated luminosities $L_{int}$= 1~ab$^{-1}$ and 10~ab$^{-1}$, respectively. 
In the two-dimensional $\chi^2$ analysis, both $\tilde{a}_\tau$ and $\tilde{d}_\tau$ couplings are assumed to deviate from their SM values at the same order. In the left (right) panel of the Fig.\ref{contour_3T}, we show 95\% C.L. contours for anomalous $\tilde a_\tau$ and $\tilde d_\tau$ couplings at $\sqrt{s}$= 3 (10)~TeV with integrated luminosities of 1 (10)~ab$^{-1}$ for the Muon collider options without and with 10\% systematic uncertainty. As we can see from the Fig.~3, the best limits without systematic uncertainty for $\tilde a_\tau$ and $\tilde d_\tau$ are $[-8.69\times10^{-4};8.76\times10^{-4}]$ ($[-2.63\times10^{-4};2. 65\times10^{-4}]$) and $[-4.85\times10^{-18};4.85\times10^{-18}]$($[-1.47\times10^{-18};1.47\times10^{-18}]$) $e$cm for the $\sqrt{s}$= 3 (10)~TeV Muon collider option. Our obtained sensitivities on the anomalous couplings without the systematic uncertanity for the process $\mu^{+}\mu^{-}  \rightarrow \tau^+\bar{\tau}^-$ with $\sqrt{s}$= 10~TeV and integrated luminosities of 10~ab$^{-1}$ can be nearly 1.5 times better than the best sensitivity derived from $\tau^+\bar{\tau}^-$
production at the Muon collider with $\sqrt{s}$= 3~TeV and integrated luminosities of 1~ab$^{-1}$. 
Moreover, the obtained sensitivity  on $\tilde{a}_\tau$ coupling for the high center-of-mass-energy Muon collider option is almost 4 times better than the current experimental limits obtained from CMS detector in the LHC at a center-of-mass energy of 13 TeV with an integrated luminosity of 138~fb$^{-1}$ via 
the production of tau lepton pairs by photon-photon fusion \cite{CMS1} while the sensitivity on $\tilde d_\tau$ improve by a factor of 20. Furthermore, considering 10 \% systematic uncertainty from possible experimental sources, the constraint on the anomalous tau lepton couplings $\tilde{a}_\tau$ and $\tilde{d}_\tau$ at high center-of-mass-energy Muon collider option with $L_{int}$= 10~ab$^{-1}$ are $[-2.71\times10^{-4};2.74\times10^{-4}]$ and $[-1.52\times10^{-18};1.52\times10^{-18}]$$e$cm, respectively. Our best limits on the anomalous $\tilde{a}_\tau$ and $\tilde{d}_\tau$ couplings at two options of the Muon colliders even with 10\% systematic uncertainty substantially improve limits reported by the DELPHI and Belle Collaborations.


\section{Conclusions}\label{sec_c}

Although the magnetic dipole moments of the electron and the muon can be precisely defined using spin precision experiments, measuring this quantity for the tau lepton is more challenging due to its much shorter lifetime than other leptons. Therefore, collider experiments are needed to study the anomalous moments of the tau lepton.
Muon colliders are crucial for particle physics because of their clean collision environment, reduced background interactions and precise energy. These colliders provide a controlled environment for high-accuracy studies of fundamental physics phenomena. In fact, beyond the LHC, such a collider is essential for studying the Higgs boson, the electroweak symmetry breaking mechanism, performing precision tests of the SM and exploring new dynamics. Therefore, possible non-standard $\tau\bar\tau \gamma$ interactions at the Muon colliders are investigated using a model-independent way an effective Lagrangian approach in this study. 
This approach is parameterised by high-dimensional operators that lead to anomalous $\tau\bar\tau \gamma$ interactions. Since non-standard $\tau\bar\tau \gamma$ couplings defined via effective Lagrangian have dimension-six, they have very strong energy dependencies. Consequently, the anomalous cross sections that include a $\tau\bar\tau \gamma$ vertex have a higher energy than the SM cross section. In this context, a potential deviation from the SM cross section of any process involving the anomalous couplings may be indicative of the existence of new physics.
For this purpose, we have examined on the phenomenological aspects of the anomalous $\tau\bar\tau \gamma$ couplings with the process $\mu^+\mu^- \to \tau^+\tau^-$ at the Muon colliders and obtain $95\%$ C. L. limits on the anomalous $\tilde{a}_\tau$ and $\tilde{d}_\tau$ parameters by using $\chi^2$
method. We conclude that Muon collider offers an ideal platform for the studying of the anomalous $\tilde{a}_\tau$ and $\tilde{d}_\tau$ couplings and 10~TeV center of mass energy with 10~ab$^{-1}$ integrated luminosity leads to best limits on them. 

\end{document}